# High-Speed Non-Volatile Barium Titanate Field Programmable Photonic Gate Array

Cristina Catalá-Lahoz[1*], Jose Roberto Rausell-Campo[1], Daniel Pérez-López[2], Lucas Güniat[3], Clarissa Convertino[3], Felix Eltes[3], Jean Fompeyrine[3], Charis Mesaritakis[4], Adonis Bogris[4], Quentin Wilmart[5], Jonathan Faugier-Tovar[5], Benoit Charbonnier[5] and José Capmany[1,2*]

[1]Photonics Research Labs, iTEAM Research Institute, Universitat Politècnica de València, Camí de Vera s/n, Valencia, 46022, Valencia, Spain.
[2]iPronics Programmable Photonics S.L., Valencia, Spain
[3]Lumiphase AG, Laubisrütistrasse 44, 8712 Stäfa, Switzerland
[4]University of West Attica, Athens, Greece
[5]CEA-Leti, Grenoble, France

*Corresponding author(s). Email: ccatala@iteam.upv.es, jcampany@iteam.upv.es

**Abstract**

Programmable integrated photonics aims to replicate the versatility of field-programmable gate arrays in the optical domain. However, scaling these systems has been prevented by the high power consumption and thermal crosstalk of conventional volatile phase shifters. Here, we demonstrate the first non-volatile field-programmable photonic gate array, implemented on a hybrid silicon-barium titanate platform. Unlike traditional thermo-optic devices that require constant power to maintain a state, our device utilizes ferroelectric domain switching to provide non-volatile memory, allowing optical circuits to be programmed and retained without any holding power or electrical bias. The hexagonal waveguide mesh integrates 58 programmable unit cells and 116 actuators, achieving nanosecond-scale switching speeds of 80 nanoseconds while reducing static power consumption to negligible levels (560 nanowatts per π phase shift). To validate this platform, we configured the mesh to perform diverse signal processing functions, including tunable filtering, 4×4 linear unitary transformations, and optical routing. This work establishes non-volatile ferroelectric silicon photonics as a scalable, heat-free platform essential for the next generation of energy-efficient photonic computing.

**Keywords:** Programmable Integrated Photonics, Non-volatile, Barium Titanate, Ferroelectric Domain Switching, Energy-efficient Photonics.

## 1. Introduction

Programmable integrated photonics (PIP)[1] promises to bring the versatility of reconfigurable electronics into photonic circuits[2], enabling flexible, low-power, and cost-effective platforms for a myriad of applications including 6G communications[3], data center interconnections[4], the Internet of Things (IoT)[5], high-tech medicine[6], artificial intelligence[7,8], photonic processing[9–11] and photonic[12–14], neuromorphic[15–19] and quantum computing[20–23]. To cope with the ever-growing requirements of these application fields there will be a need for scaling programmable photonic circuits to integrate $10^4$ actuators and beyond in the forthcoming years and this will require the simultaneous addressing of several challenges[24,25] that so far have remained elusive to the current state of the art silicon photonics thermo-optic (TO) and micro-electromechanical (MEMS) phase shifting technologies[26–29]. In essence, programmable unit cells (PUCs) with minimal footprint (below 100 μm), loss (below 0.15 dB), power consumption (below 1 mW/π) and featuring tuning speeds in the nanosecond scale, will be required to implement complex multiport interferometers and waveguide meshes. TO-based phase shifters incur moderate static power consumption (1-10 mW/π), while those based on mechanical effects require a constant bias when the device is in the standby state. Hence, the static energy consumption can easily scale with the number of actuators, leading to a significant increase in the power budget of the programmable processor.

Non-volatile materials such as phase-change compounds, ferroelectrics, and memristive oxides have been proposed as alternatives for the implementation of integrated phase shifters, since their optical state can be maintained without any power consumption or bias. Among the different options, phase-change compounds, ferroelectrics, and memristive oxides have been researched during the last years as alternatives, offering reversible switching with ultralow power and potentially fast response[30]. While each of these material platforms has shown promising advances, their respective limitations have restricted experimental demonstrations to only modestly sized circuits, far below the scale required for fully programmable, general-purpose photonic processors.

Phase-change materials (PCMs), such as $Ge_2Sb_2Se_5$ (GST), are chalcogenides that can be electrically switched between bistable microstructural states of ordered crystalline and disordered amorphous states, providing a phase shift. Integration of PCMs in a silicon waveguide has been reported, where the phase shifting is achieved via the TO effect[31–33], and the technology shows potential fast switching speed (0.1 μs, 1 mW) for the setup of the amorphization phase and potential low footprint ($L_\pi$=10-50 μm). However, the switching speed for transition to the crystallization phase is of the order of milliseconds, leading to power consumption in the tens of mW. Another disadvantage stems from the high losses (>0.33 dB) arising from the proximity of the PCMs to the optical mode in the silicon waveguide. These limitations hamper the scalability of photonic processors based on this technology.

Memristors typically consist of a conductor-insulator-conductor tri-layer[34]. The conductor layers can be made of metals or semiconductors, whereas the insulator is a dielectric material. The application of enough voltage forms a conduction filament that switches the memristor from an initial high resistance (HRS) into a low resistance state (LRS). The memristor can be returned to the HRS by reversing the bias polarity, which breaks the filaments. The combination of this mechanism with the free carrier dispersion effect is then exploited to provide a controllable phase shift. A recent experiment[35] reported a silicon photonic phase shifter based on the heterogeneous integration of $GaAs/Al_2O_3/Si$ memristor in a 10 μm microring resonator featuring low insertion loss (<0.05 dB). While results in terms of switching speed (< 1 ns), switching powers (150-360 μW), and insertion losses (0.27 dB) are promising, the compactness of phase shifters is limited by the large value of $L_\pi$ (350 μm).

Ferroelectrics and, in particular, barium titanate (BTO), have recently emerged as a promising material for efficient phase shifters. BTO features one of the highest Pockels coefficients reported ($r_{42}$ of ~923 pm V$^{-1}$), and its potential for large-scale integration with silicon photonics circuits has been recently demonstrated through a combination of epitaxial growth and direct wafer bonding[36]. BTO also allows the non-volatile storage of optical information by switching non-volatile ferroelectric domains through the application of an electric field. The first demonstration did result in multilevel storage over 8 levels[37], although initial devices developed in this platform showed large $L_\pi$ (1mm) and moderate switching speed (~microsecond) for non-volatile operation. This platform has a clear potential for improvement to obtain compact footprints, ultralow insertion loss, sub-milliwatt switching, nanosecond reconfigurability, and non-volatility.

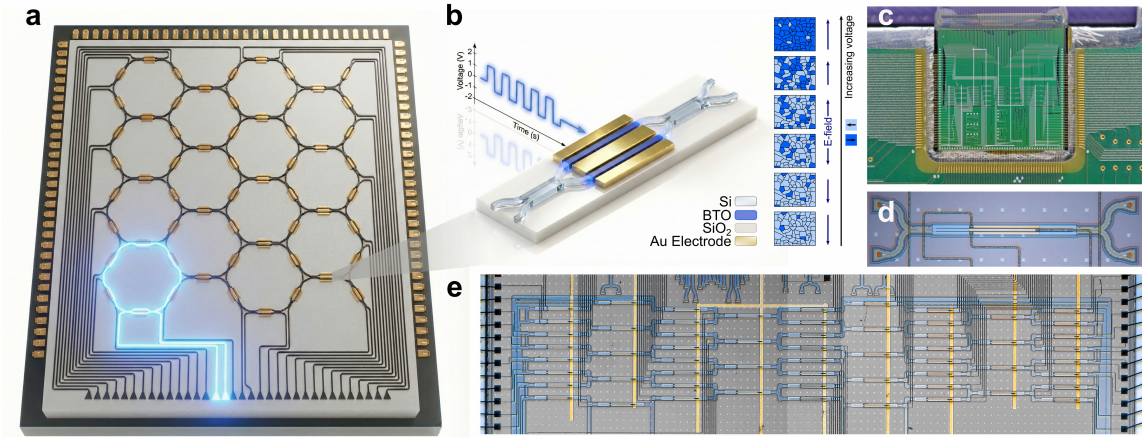

**Fig. 1. Design principle and physical realization of the non-volatile FPPGA. a**, Artistic rendering of the programmable photonic chip based on a hexagonal waveguide mesh topology. **b**, Illustration of the non-volatile phase shifting mechanism on the hybrid Si-BTO platform. Voltage pulses reorient the ferroelectric domains (blue), modifying the effective refractive index without static power consumption. **c**, Photograph of the fabricated and wire-bonded chip mounted on a PCB. **d**, Optical microscope image showing a detail of a single NV-PUC. **e**, Zoomed-out microscope image of the fabricated hexagonal mesh, displaying the large-scale integration of unit cells and interconnected BTO actuators.

Here we present, to our knowledge, the first non-volatile, field-programmable photonic gate array (FPPGA) based on barium titanate ($BaTiO_3$) actuators. The FPPGA integrates 116 actuators or phase shifters in a flattened hexagonal waveguide mesh configuration and achieves nanosecond-scale switching with sub-microwatt power consumption per actuator, two orders of magnitude below the state of the art. We experimentally demonstrate several key signal processing applications, including reconfigurable filters, unitary linear transformations, optical routing, and splitting. Moreover, we show the capability of the individual programmable unit cells (PUCs) to set and retain a specific state after the external bias is turned off, thanks to their non-volatile behavior. These characteristics mark a significant step forward towards large-scale, power-efficient programmable photonic systems.

## 2. Results

Programmable photonic integrated circuits use meshes of interconnected waveguides, where reconfigurable 2×2 elements known as PUCs or gates define and control the amplitude and phase of the optical signals that propagate through the network. Following the philosophy of electronic field-programmable gate arrays (FPGAs), a photonic FPGA (FPPGA) can be reprogrammed to realize a variety of photonic circuits and linear multiport transformations by configuring its PUCs and selecting the appropriate ports[38,39]. A common implementation of a PUC is based on a balanced Mach-Zehnder interferometer (MZI) consisting of two 50:50 multimode interferometers (MMIs) and an electrical biased phase shifter on each arm. Using a PUC with non-volatile actuators (NV-PUC) as the basic building block eliminates static power consumption, since each state can be set once and retained without the need to apply a continuous bias.

Figure 1 illustrates the design principle and physical realization of the non-volatile FPPGA. The schematic in Fig. 1a depicts the FPPGA based on a hexagonal waveguide mesh topology, enabling software-defined programming of specific photonic circuits. The underlying mechanism of this reconfigurability, detailed in Fig. 1b, stems from the implementation of PUCs on a hybrid Silicon-Barium Titanate (Si-BTO) platform. By applying voltage pulses, the ferroelectric domains within the BTO layer are reoriented, altering the effective refractive index and providing non-volatile phase shifting. The realization of this design is presented in Fig. 1c, which shows the fabricated and wire-bonded chip on a printed circuit board. Optical microscope images provide further detail on the device specifics: Fig. 1d displays a close-up of a single programmable unit cell, while Fig. 1e illustrates the large-scale integration of these cells into the interconnected hexagonal mesh with a flattened design[40]. The footprint of the fabricated circuit is 2×10 mm² and includes 58 PUCs, 116 BTO phase shifters, and 34 optical input/output ports.

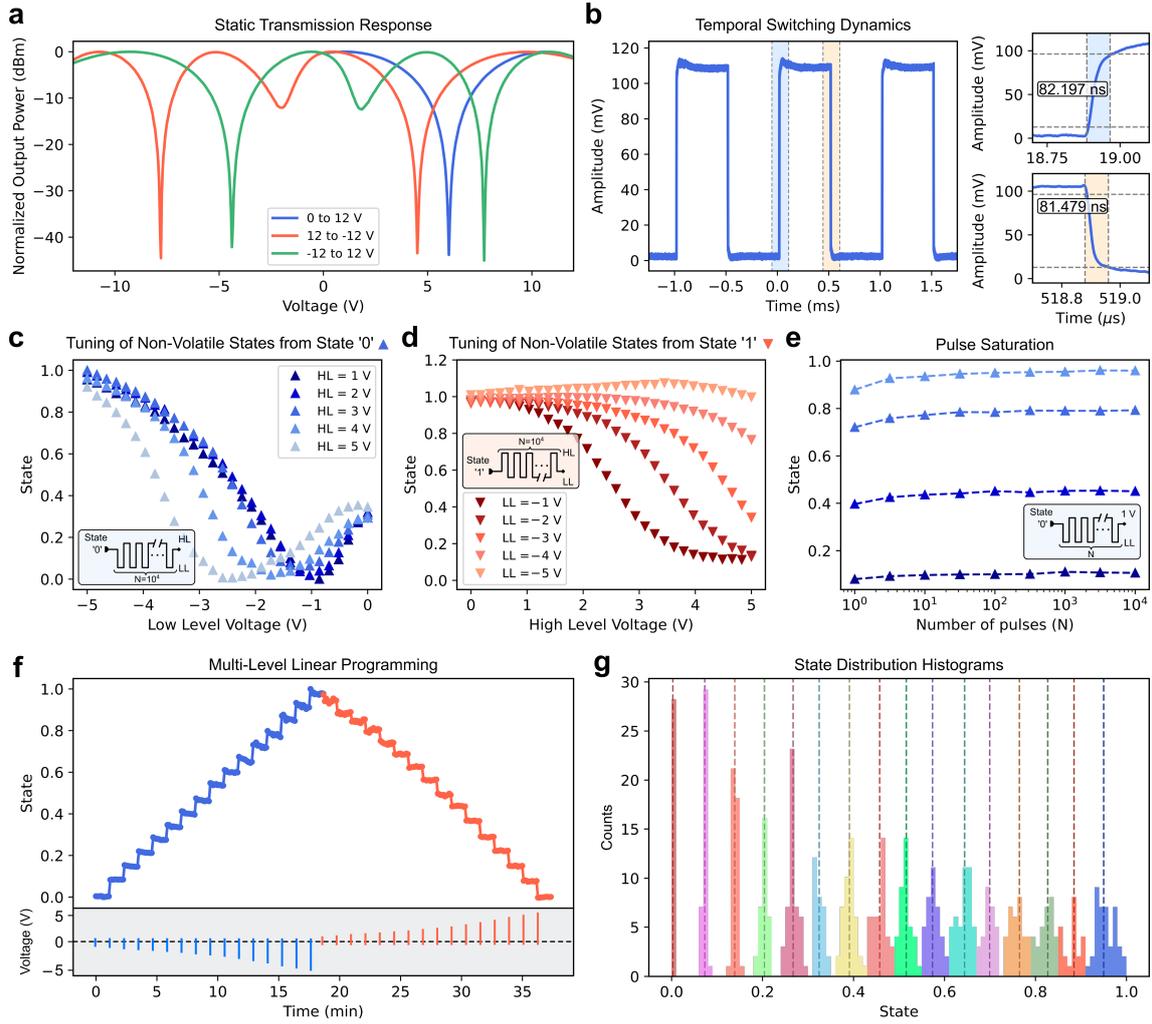

**Fig. 2. Characterization of the NV-PUC.** **a**, Static transmission response of the MZI showing the characteristic butterfly-shaped hysteresis loop resulting from ferroelectric domain reorientation under DC voltage sweeps. **b**, Temporal switching dynamics demonstrating symmetric rise and fall times of ~82 ns, driven by the Pockels effect. **c**, **d**, Tuning of non-volatile states starting from State '0' (c) and State '1' (d) by applying pulse trains with varying Low Level (LL) and High Level (HL) voltages. **e**, State saturation behavior as a function of the number of applied pulses (N), showing stability beyond $10^2$ pulses. **f**, Multi-level linear programming demonstrating a stable "staircase" response with 16 distinct phase levels. **g**, State distribution histograms obtained from 600 random transitions, validating the accuracy and reproducibility of the 16 programmable states.

### 2.1. Non-volatile programmable unit cell

The NV-PUC serves as the basic reconfigurable element of the mesh circuit. Phase shifters are implemented using ferroelectric BTO integrated with silicon waveguides, following the fabrication process described in the Methods section. Programming of the unit cell relies on electrically controlling the BTO actuators applying a transverse electric field. By adjusting the differential phase between the interferometer arms, the NV-PUC can switch between its bar and cross states or provide analog coupling tuning.

The integration of BTO actuators enables the NV-PUC to operate in both volatile and non-volatile modes depending on the characteristics of the applied voltage. When the NV-PUC is operated in volatile mode, a DC voltage modifies the refractive index of the BTO through the Pockels effect. When operated in non-volatile regime, the change in the configuration of its ferroelectric domains, produces a linear and reversible phase change. The combination of both effects creates a butterfly shaped hysteresis loop in the effective refractive index. The history-dependent optical response can be experimentally observed in the interferometric curves of the PUCs. When the voltage sweep begins at 0 V, where only one domain configuration is stable, we observe the response of a standard MZI. However, when the sweep starts from a previously biased state, either positive or negative, the BTO film presents a range of possible refractive index configurations. As the voltage is swept, part of the ferroelectric domain population begins to reorient. When the direction of the applied electric field is reversed, the hysteresis curve of the refractive index is transferred to the interferometric response creating a similar butterfly-shape. Figure 2a shows the MZI response when we apply a DC voltage, sweeping from 0 to 12 V and from +12 (-12) to -12 (+12) V. The interplay between the instantaneous Pockels effect and the slower domain reorientation produces the characteristic butterfly-shaped hysteresis observed when sweeping from positive to negative voltages or vice versa.

Building on this static characterization, the BTO phase shifters result in an ultra-low static power consumption of just 560nW/$\pi$ corresponding to a $V_\pi = 5$ V. This represents an outstanding improvement for the scalability of programmable photonics, where standard thermo-optic heaters typically require 1.3 – 20 mW per phase shift, leading to significant thermal management challenges in large-scale circuits. Another significant performance advantage of the BTO platform is observed in its temporal dynamics. Figure 2b illustrates the optical response of the MZI driven by 2 MHz square pulses with 4.8 $V_{pp}$ and 1.2 V of bias. The device demonstrates highly symmetric switching with rise and fall times around 82 ns (measured from 10% to 90%). This nanosecond-scale response is intrinsic to the electro-optic nature of the Pockels effect, offering a speed advantage of 3-4 orders of magnitude compared to conventional thermo-optic phase shifters (2 - 100 μs), which are physically limited by slow thermal diffusion and are also sensitive to thermal crosstalk[3,26]. The fabricated NV-PUC exhibits an insertion loss of 1.48 dB and a basic delay unit of 12.73 ps, with a total length of 950 μm, being 350 μm the length of the phase shifters and the rest being the MMIs and bends.

In non-volatile operation, the NV-PUC takes advantage of the fact that the ferroelectric domain configuration in BTO can be permanently modified, which changes the effective Pockels coefficient and therefore, the refractive index at a given applied bias. For any non-zero voltage, the material can support multiple stable refractive-index states depending on how the domains are distributed. To program these states, we apply short electrical pulses (on the order of nanoseconds to microseconds) at the desired bias level and voltage. These pulses are shorter than the characteristic domain switching time of the BTO film, so instead of driving the material into a fully poled state, they produce controlled mixed-polarization domain configurations. Adjusting the bias and amplitude voltages (or low and high voltage amplitudes/levels) at which the pulses are delivered allows access to different domain mixtures, each corresponding to a distinct refractive-index state that remains stable after the pulse sequence is removed.

To overcome the path-dependence inherent to ferroelectric domains, we implemented a reset protocol to ensure identical initial conditions for every measurement. We established two different boundary states, designated as State '0' and State '1', which frame the accessible operation range. These states are prepared via a two-stage sequence: first, the previous domain history is erased using a decaying alternating electric field; second, the polarization is saturated using a train of $10^4$ initialization pulses at +5 V (for State '0') or -5 V (for State '1'), with a pulse duration of 300 ns. Without this preconditioning, the device would settle into a random intermediate state determined by its prior usage. Once the device is initialized, the ferroelectric polarization saturates, meaning subsequent pulses of the same polarity induce no further refractive index modification. Consequently, tuning is achieved by applying $10^4$ pulses of the opposite polarity while maintaining a bias matching the initialization sign (see Supplementary Note 1 for details on the procedure for erasing and setting the states). Figure 2c characterizes the tuning dynamics starting from State '0'. In this regime, while holding a positive High Level (HL) bias between 1 and 5 V, we apply a pulse train that sweeps the Low Level (LL) voltage from 0 down to -5 V. The complementary behavior is shown in Figure 2d, which illustrates tuning from State '1'. Here, the device is tuned by sweeping the HL voltage from 0 to 5 V against distinct negative LL baselines. Under this optimized driving scheme, a phase shifter with an active length of just 350 µm achieves a full non-volatile π phase shift with an extinction ratio of 12 dB.

We also investigated the device's non-volatile behavior as a function of the number of applied pulses (N). To do this, we applied consecutive 300 ns pulses of constant amplitude while varying N. Figure 2e shows that for intermediate states (between 0 and 1), the achieved state saturates beyond $N = 10^2$ pulses. This means that to establish a state in the device, the total required pulse train time would be 60 µs.

The ability to retain multiple intermediate states allows this device to act as a tunable building block for programmable photonics. To demonstrate this accurate control, we initialized the NV-PUC to State '0' and applied a sequence of specific pulses. The amplitudes of these pulses were pre-calculated using the data from Fig. 2c–d to ensure that the optical steps were linear and equally spaced. We programmed sixteen distinct states, holding each one for 1.2 minutes to verify its stability before moving to the next. The sequence was then reversed using pulses of the opposite polarity. As shown in Fig. 2f, the device produced a stable, linear 'staircase' response, successfully resolving sixteen equally spaced states in both directions. To ensure that any state can be set reliably regardless of the previous value, we performed a random switching test involving 600 transitions. In this experiment, we randomly selected target states from the sixteen available levels. To guarantee accuracy, we reset the device to State '0' before every new setting. The resulting performance is shown in Fig. 2g. The histogram displays sixteen clearly separated peaks, proving that the device creates consistent and reproducible states.

## 2.2. Hexagonal mesh programmable circuit

We initiated the study with the static characterization of the hexagonal mesh by applying DC voltage sweeps to the constituent PUCs to generate their optical response curves. The sequence of characterization was defined using a custom graph-based algorithm (see supplementary Note 2 for setup and characterization algorithm details). The resulting parameters were subsequently stored in a database, allowing us to precisely configure the specific coupling ratios required for later experiments. Following system calibration, we evaluated the spectral operating range of the input and output grating couplers. Our measurements indicate that the device achieves peak efficiency in the 1565 ± 15 nm window.

The hexagonal waveguide mesh can be configured to emulate the hardware architecture of a diverse array of optical circuits, functioning as a general-purpose photonic processor. The mesh supports, among other functions, traditional feedforward and feedback finite impulse response (FIR) and infinite impulse response (IIR) filters, as well as complex multiple-input/multiple-output (MIMO) linear optical transformers.

The unbalanced Mach-Zehnder interferometer (UMZI) works as a periodic notch filter. These devices are fundamental components in the construction of lattice filters and FIR transversal filters. By manipulating the PUC settings, we successfully synthesized UMZI structures with path length differences equivalent to 2 and 4 times the standard PUC length. The experimental data for these configurations are displayed in Figure 3a. These path imbalances resulted in free spectral ranges (FSR) of 43.65 GHz and 19.02 GHz, respectively.

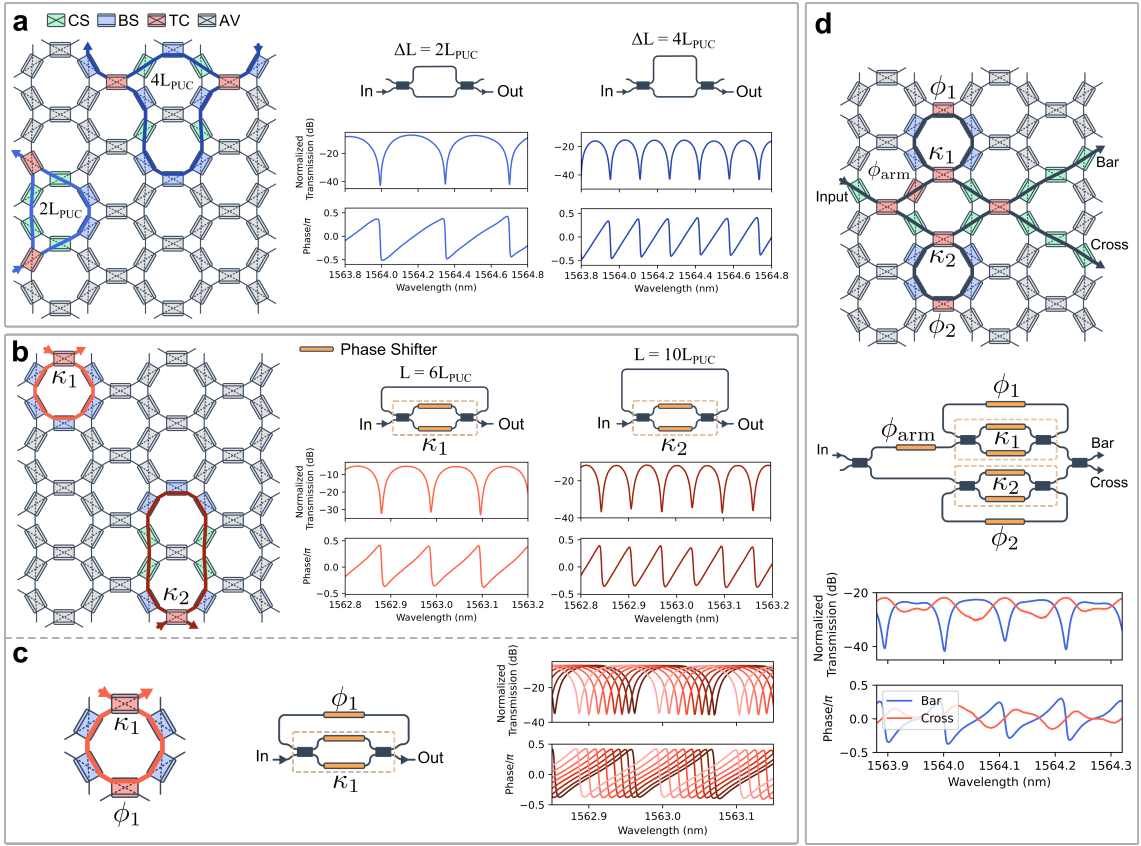

**Fig. 3. Implementation of reconfigurable optical filters on the hexagonal mesh. a**, Synthesis of UMZIs with path length differences of $2L_{PUC}$ and $4L_{PUC}$, showing the measured normalized transmission spectra and phase response. **b**, Optical ring resonators configured with cavity perimeters of $6L_{PUC}$ and $10L_{PUC}$. **c**, Demonstration of IIR filter tunability achieved by independent control of the PUC phase shifters to sweep the resonance peak. **d**, Configuration and measured spectral response of a complex two-order RAMZI filter.

We also implemented ring cavities, which operate as periodic filters. Using these structures, we can implement all-pole IIR notch filters, and combining them with other structures, they can realize both IIR notch and hybrid FIR + IIR bandpass filters. These resonators are essential for assembling complex architectures, such as Coupled Resonator Optical Waveguides (CROW) or Ring-Assisted Mach-Zehnder Interferometers (RAMZI). Through precise tuning of the unit cells, we realized single optical ring resonators with cavity perimeters of 6 and 10 PUC lengths. Figure 3b illustrates the measurements for these cavities, which exhibited FSR values of 13.27 GHz and 7.95 GHz, respectively. Figure 3c demonstrates the tunability of the IIR filter. This is obtained by tuning both actuators of the PUC acting as a phase shifter which allows the control of the coupling and output phase independently. By varying the voltage injected into both phase shifters, the resonance peak can be swept across a complete spectral period.

Leveraging this flexibility, we built more complex filters, such as the two-ring RAMZI filter depicted in Figure 3d. The measurement results for this structure display both the bar and cross output ports, which are complementary and feature an FSR of 13.14 GHz (see Supplementary Note 3 for additional filter measurements).

Beyond filtering applications, we leveraged the hexagonal waveguide mesh to implement versatile MIMO linear optical transformations. The mesh's programmability allows it to emulate various connectivity schemes, ranging from unitary matrix operations to complex routing and broadcasting networks.

We programmed the hexagonal mesh to implement a rectangular unitary architecture, creating a 4×4 programmable linear optical transformer. By defining six active tunable couplers (S1–S6), as shown in Fig. 4a, we established a reconfigurable switch capable of executing various unitary operations. We synthesized three distinct transfer matrices (M) to demonstrate the device's switching capabilities. The corresponding bar graphs display the normalized output power for each configuration. The device exhibited exceptional fidelity, with normalized output powers consistently exceeding 0.98 and a high transmission ratio between desired and undesired connections (~25 dB).

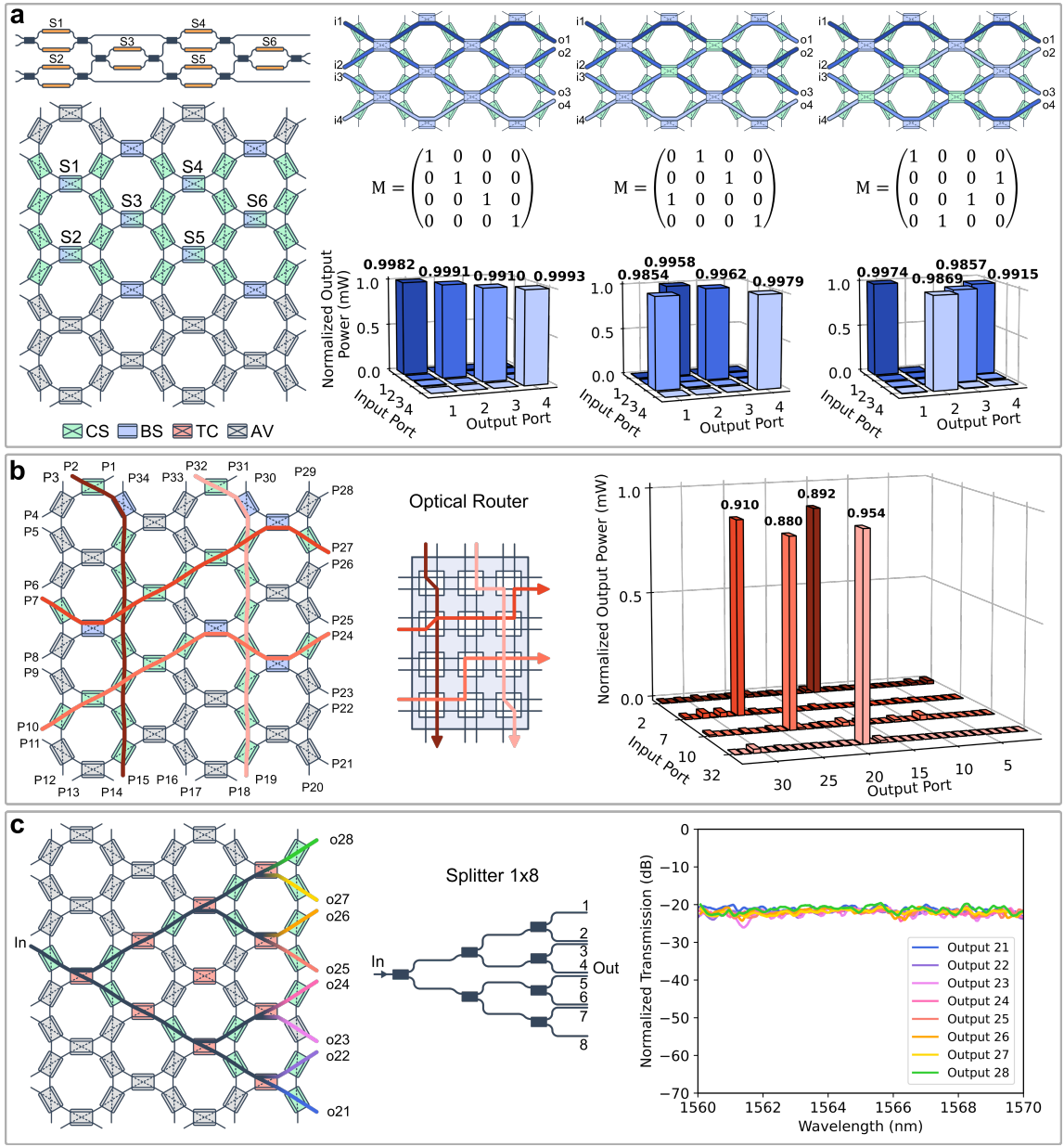

**Fig. 4. Linear transformations and routing on the photonic processor. a**, 4×4 programmable linear optical transformer. Left: Mesh schematic defining six active tunable couplers (S1–S6). Right: Target unitary transfer matrices and measured normalized output powers for three different configurations, demonstrating high fidelity. **b**, Configuration of the mesh as an optical router, allowing the simultaneous routing of four independent signals without interference. The bar chart displays the normalized output power for each target port. **c**, Demonstration of a 1×8 optical splitter implemented via a cascading tree structure. The graph shows the superimposed transmission spectra of the eight output ports (o21–o28), indicating uniform power distribution and broadband operation.

To validate the platform's interconnect capabilities, we programmed the mesh as a flexible optical router. Figure 4b presents a multi-path scenario where four independent optical signals are routed simultaneously across the chip lattice without interference. We established four distinct trajectories: Port 2 to 15, 7 to 27, 10 to 24, and 32 to 19. The measured normalized output powers for these paths were 0.892, 0.91, 0.88, and 0.954, respectively. These high transmission values indicate that the mesh can support multiple intersecting signal flows, dynamically creating optical links between arbitrary points on the perimeter.

Finally, we demonstrated the mesh's capacity for broadcasting applications by configuring it as a 1×8 optical splitter. As shown in the schematic in Fig. 4c, a single input signal is distributed through a cascading tree structure to eight distinct output ports (o21–o28). The spectral response measurement shows a highly uniform power distribution across all eight channels in the 1560–1570 nm wavelength range. The overlapping traces indicate balanced splitting ratios and broadband operation, proving the mesh is suitable for one-to-many signal distribution tasks.

## 3. Discussion

We have demonstrated, for the first time to our knowledge, an extended programmable photonic processor capable of non-volatile operation, addressing the power wall that currently constrains the scaling of programmable integrated photonics. By integrating ferroelectric BTO actuators into a silicon photonic hexagonal waveguide mesh, we realized a platform that combines the zero-static power advantages of non-volatile memory with the high-speed performance required for dynamic signal processing. While non-volatile functionalities were recently explored in reduced scale demonstrations using PCMs[41], our work represents the first realization of a fully reconfigurable, general-purpose hexagonal FPPGA based on ferroelectric actuators.

The central achievement of this work is the decoupling of circuit complexity from energy consumption. In conventional programmable circuits based on thermo-optic phase shifters, maintaining a specific circuit configuration requires continuous electrical biasing. With typical consumption rates of 1–10 mW per actuator, a large-scale mesh comprising thousands of unit cells would consume tens of watts, necessitating complex thermal management solutions to prevent crosstalk and instability. In contrast, our BTO-based architecture consumes very low power (560 nW/π per actuator) when driven in a volatile mode. Further, if it is driven as a non-volatile, once programmed, the state is retained by the ferroelectric domain orientation without external bias. This "set-and-forget" capability suggests that future processors could scale to thousands of actuators without a corresponding increase in the static power budget.

To contextualize the performance of the ferroelectric Si-BTO platform, Table 1 benchmarks our device against the leading phase-shifting technologies currently used in programmable photonics.

**Table 1.** Performance benchmark of phase-shifting mechanisms for large-scale photonic circuits.

| Technology | Physical Mechanism | Volatility | Static Power Consumption | Switching Speed | Switching Energy | Insertion Loss | Actuator Length | Thermal Crosstalk |
|---|---|---|---|---|---|---|---|---|
| Thermo-Optic (w/o trenches)[26] | Temperature dependence | Volatile | ~25 mW/π | 2.69 μs | 67.25 nJ | 0.23 dB | 61.6 μm | High |
| Thermo-Optic (with trenches)[3] | Temperature dependence | Volatile | 1.3 mW/π | 100 μs | 130 nJ | 0.48 dB | ~220 μm | Low |
| MEMS[28,42] | Electrostatic actuation | Volatile | < 1 μW | 1.54 μs | 0.2 nJ | 0.33 dB | 50 μm | Zero |
| PCM: GST[43,44] | Phase transition (Amorph/Cryst) | Non-volatile | Zero | 100 ns and 0.6 ms | 180 pJ and 17 nJ | 1.6 dB | 5 μm | High |
| BTO (This work) | Ferroelectric | Volatile | ~560 nW/π | ~80 ns | 44.8 fJ | 1.48 dB | 350 μm | Zero |
| | | Non-volatile | Zero | ~60 μs | | | | |

As shown in Table 1, our device bridges the performance gap between existing switching technologies. Phase-change materials typically rely on transitions between amorphous and crystalline states, which can introduce significant optical absorption (>0.33 dB per element) and are limited by slower crystallization speeds (milliseconds). In contrast, our BTO platform leverages the Pockels effect, enabling switching speeds of 80 ns, several orders of magnitude faster than both PCMs and thermo-optic heaters.

However, regarding optical loss, our current prototype exhibits an insertion loss of 1.48 dB per PUC. While this is currently higher than the intrinsic loss of optimized PCM or MEMS devices, it is crucial to distinguish the source of this attenuation. Unlike PCMs, where loss is often intrinsic to the material state, the losses in our device are primarily due to the hybrid integration process. Specifically, the mode transitions (tapers) between silicon and Si-BTO waveguides. Because BTO itself is inherently transparent at telecom wavelengths, there is a clear engineering path to reduce these losses to <0.5 dB in future generations by optimizing the taper geometry and buffer layer thickness.

Despite the current insertion loss, the experimental demonstrations, ranging from tunable FIR/IIR filters to rectangular unitary architectures, validate the optical robustness of the platform. However, it is important to clarify the operational mode used in these system-level experiments. While the individual unit cells have been proven to support non-volatile memory storage, the complex mesh configurations presented here were characterized using volatile actuation. Controlling the entire mesh in non-volatile mode requires an electronic interface capable of distributing specific programming pulses to all electrodes independently. In our current experimental setup, which relies on a single pulse source, such addressing is limited to manual operation. Consequently, the development of an integrated electronic switching matrix to automate pulse distribution is a critical next step to fully unlock the platform's zero-static power potential in large-scale circuits.

Finally, the reliance on the Pockels effect fundamentally improves the thermal management strategy compared to conventional platforms. Unlike thermo-optic phase shifters, which function by locally heating the waveguide and inherently suffer from thermal crosstalk between adjacent components, our ferroelectric actuators are electric-field driven and generate no waste heat. This effectively eliminates intra-device thermal crosstalk, a major bottleneck in dense photonic integration. However, the ferroelectric material remains sensitive to global ambient temperature fluctuations (see Supplementary Note 4 for characterization of the device's temperature dependence). Therefore, while the system is free from self-induced heating instabilities, precise stabilization of the chip substrate using a Thermoelectric Cooler (TEC) is required to maintain performance against external environmental drift.

In conclusion, the realization of a non-volatile, field-programmable waveguide mesh fundamentally alters the scaling prospects of integrated photonics. By eliminating the static power consumption and thermal crosstalk inherent to thermo-optic approaches, this platform overcomes the primary bottlenecks preventing the development of large-scale optical processors. While future efforts must address insertion loss and electronic co-integration, the results presented here provide a clear blueprint for sustainable, high-speed photonic computing. This technology is poised to play a critical role in next-generation hardware, where energy efficiency and adaptability are paramount.

# 4. Methods

## 4.1. Fabrication

The 200 mm base wafers used for the BTO integration process were fabricated at CEA-Leti using a 220 nm silicon-on-insulator (SOI) platform with a 3 µm buried oxide (BOX) layer. To prepare a surface suitable for BTO bonding, we targeted a top oxide thickness of 350 nm, achieved through sequential oxide deposition and chemical-mechanical polishing (CMP) steps. Final measurements confirmed a resulting thickness of approximately 365 nm (Supplementary Note 5). Subsequently, the wafers were transferred to Lumiphase, which transfers the BTO layer. For the interlayer coupling between Si passives and BTO phase shifters, a taper of the silicon waveguide was designed to transfer the mode to the Si-BTO waveguide. Finally, waveguide cladding and metallization through vias complete the fabrication process.

## 4.1. Characterization and measurements

The characterization of the NV-PUC was performed using test structures located on the same chip as the hexagonal mesh. The initial experimental setup consists of a tunable CW laser (EXFO T100S-HP) followed by a polarization controller. Light is coupled into and out of the PUC's grating couplers using two cleaved fibers mounted on 8-degree tilted wedges with Tapered V-Groove Fiber Holders. At the output of the PUC, the detected optical power is measured using a programmable power meter (EXFO FTBx-1750). To control the PUC actuators, RF probes with a 250 µm pitch were used to apply two different types of signals from a wavefunction generator (Teledyne T3AFG120). These signals are used to set and erase the various states of the PUC.

The procedure for setting a state begins by erasing the ferroelectric domain state. To achieve this, a sine signal of 1.3 MHz, amplitude-modulated with a down-ramp waveform (33.33 Hz), is applied for 2 seconds to reorganize the domains into an arbitrary distribution. Subsequently, two specific ferroelectric domain states (0 and 1) were defined by applying a sequence of $10^4$ pulses, each with a duration of 300 ns at 5 V or -5 V (depending on the state 0 or 1). This pulse combination results in a saturated state for both polling directions. After setting the initial state (0 or 1), a train of N pulses (300 ns duration) is applied with specific upper and lower amplitudes ($V_{HL}$ and $V_{LL}$). Once the state is set, the voltage source is turned off, and the optical output power is measured (Supplementary Note 1). The erase signal and the pulse train for setting the states were generated using the same waveform generator, one per channel of the device. An electrical amplifier was placed at the output of the erase signal, and its output, along with the second channel (containing the pulse train), was connected to a coaxial switch (Radiall R570412000LP) used to sequentially apply the two signals.

For the mesh measurements, the device was first automatically characterized. To do this, a bank of power meters (EXFO LTB-12 & FTBx-1750) was employed to monitor the optical power at all mesh output ports. This automated characterization routine relies on a graph representation that replicates the mesh topology. We selected a single fixed input port and, using the graph, determined the shortest paths to the remaining ports. By optimizing the output power for each path, we sequentially characterized the individual MZIs along that path, iterating the process for all other ports (Supplementary Note 2). Finally, a passive component tester (EXFO CT440) was used to acquire transmission spectra around 1560 nm with a spectral resolution of 1 pm.

## Data availability

The data that support the findings of this study are available from the corresponding authors upon reasonable request.

## Code availability

The code that supports the plots and data analysis of this study is available from the corresponding authors upon reasonable request.

## Declarations

The authors declare no conflicts of interest.

## Supplementary information

See the attachment file.

## Author contributions

**J.C.** and **D.P.L.** conceived the concept and supervised the overall project. **J.R.R.C.** designed the layout of the hexagonal mesh. **Q.W., J.F.T.** and **B.C.** were responsible for the silicon wafer fabrication and planarization processes. **F.E., L.G., C.C.** and **J.F.** carried out the BTO integration. **C.C.L.** developed the control software, performed the characterization of the programmable mesh and carried out the experimental demonstration of the applications. **C.C.L.** and **J.R.R.C.** performed the measurements of the non-volatile test structures. **J.C.** wrote the introduction and **C.C.L.** wrote the original draft of the manuscript. **C.M.** and **A.B.** participated in the discussion of the results and provided comments on the manuscript. All authors reviewed and approved the final version.

## Acknowledgements

This work was supported by the European Commission through projects H2020-ICT2019-2 Neoteric 871330 Project, Horizon Digital Emerging-01 101070195 Prometheus. J.C and D.P.L. acknowledges funding through ERC Grants ERC-AdG-2022-ANBIT-101097092, ERC-PoC-2025-TRANSBIT-101241773 and ERC-StG-2022-LS-Photonics-101076175.

# Supplementary Information
High-Speed Non-Volatile Barium Titanate Field Programmable Photonic Gate Array

Cristina Catalá-Lahoz[1*], Jose Roberto Rausell-Campo[1], Daniel Pérez-López[2], Lucas Güniat[3], Clarissa Convertino[3], Felix Eltes[3], Jean Fompeyrine[3], Charis Mesaritakis[4], Adonis Bogris[4], Quentin Wilmart[5], Jonathan Faugier-Tovar[5], Benoit Charbonnier[5] and José Capmany[1,2*]

[1]Photonics Research Labs, iTEAM Research Institute, Universitat Politècnica de València, Camí de Vera s/n, Valencia, 46022, Valencia, Spain.
[2]iPronics Programmable Photonics S.L., Valencia, Spain
[3]Lumiphase AG, Laubisrütistrasse 44, 8712 Stäfa, Switzerland
[4]University of West Attica, Athens, Greece
[5]CEA-Leti, Grenoble, France

*Corresponding author(s). Email: ccatala@iteam.upv.es, jcampany@iteam.upv.es

# Contents

**Supplementary Note 1.** Electrical driving schemes for non-volatile operation.

**Optical Characterization Setup**

The characterization of the Non-Volatile Programmable Unit Cell (NV-PUC) was conducted using dedicated test structures fabricated on the same chip as the hexagonal mesh. The optical measurement setup utilized a tunable Continuous Wave (CW) laser (EXFO T100S-HP) acting as the light source. To ensure optimal coupling efficiency, the polarization of the input light was adjusted using a manual polarization controller prior to coupling.

Light coupling into and out of the PUC's grating couplers was achieved using two cleaved fibers mounted on 8º tilted wedges equipped with Tapered V-Groove Fiber Holders. The optical power transmitted through the device was detected and analyzed using a programmable power meter (EXFO FTBx-1750).

**Electrical Control**

Electrical actuation of the PUC was performed using Radio Frequency (RF) probes with a 250 µm pitch. These probes delivered control signals generated by a dual-channel arbitrary waveform generator (Teledyne T3AFG120). The signal routing was designed to switch sequentially between an "Erase" signal and a "Write/Set" signal without manual intervention:

- Channel 1 (Erase Line): Connected to an electrical amplifier to achieve the necessary voltage compliance for domain reorganization.
- Channel 2 (Pulse Train/Write Line): Connected directly to the output switch.

Both channels were routed into a coaxial switch (Radiall R570412000LP), allowing for the precise sequential application of the erasure and setting waveforms to the PUC. Figure S1 shows a schematic of the complete setup.

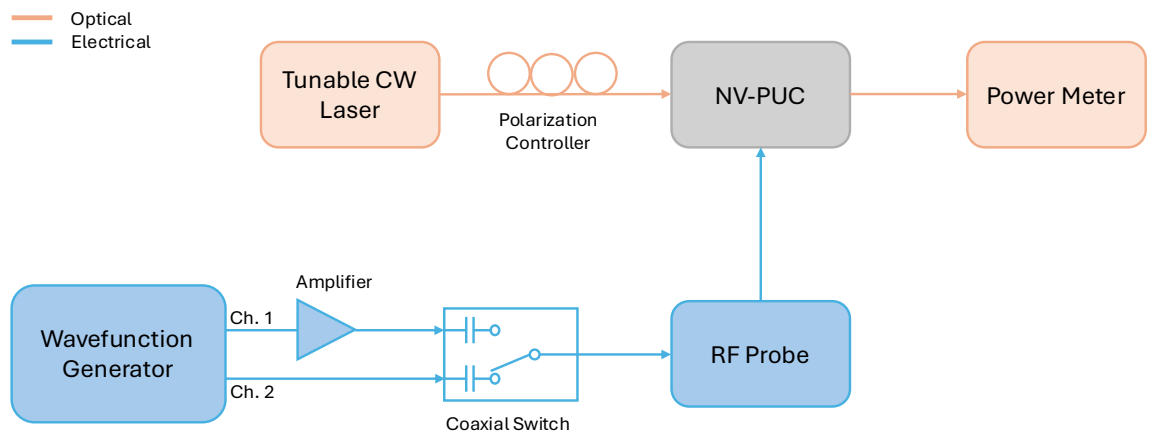

**Fig. S1. Schematic of the NV-PUC Characterization Setup.** The optical path consists of a tunable CW laser, a polarization controller, and a power meter. The electrical actuation is managed by a dual-channel wavefunction generator; Channel 1 (Erase) is amplified, while Channel 2 (Write) is connected directly. A coaxial switch routes the appropriate signal to the RF probe for domain erasure or state setting.

**Experimental Procedure**

To accurately characterize the ferroelectric switching dynamics, a four-step protocol was defined[1]:

**Step I: Domain Erasure (Reset).** To eliminate hysteresis and history effects from previous measurements, the ferroelectric domain state was first erased. A sine wave with a frequency of 1.3 MHz was applied for a duration of 2 seconds. This signal was amplitude-modulated with a down-ramp waveform (33.33 Hz) to gradually reduce the field strength, reorganizing the domains into an arbitrary, randomized distribution.

**Step II: Saturation (poling).** Subsequently, two specific ferroelectric domain states (State 0 and State 1) were defined by applying a sequence of $10^4$ pulses.

- Pulse Duration: 300 ns.
- Amplitude: +5 V for State 0, -5 V for State 1.

This pulse combination results in a saturated state for both poling directions.

**Step III: Set a state.** After setting the initial saturated state (0 or 1), we apply a train of N pulses, each with a duration of 300 ns (with 300 ns separation as well). These pulses alternated between specific high level ($V_{HL}$) and low level ($V_{LL}$) amplitudes to induce partial switching.

**Step IV: Measure.** Immediately following the pulse train, the voltage source is turned off. The optical transmission is then measured to quantify the non-volatile change in the PUC state.

The temporal evolution of the applied voltage signals throughout the four phases is depicted in Fig. S2.

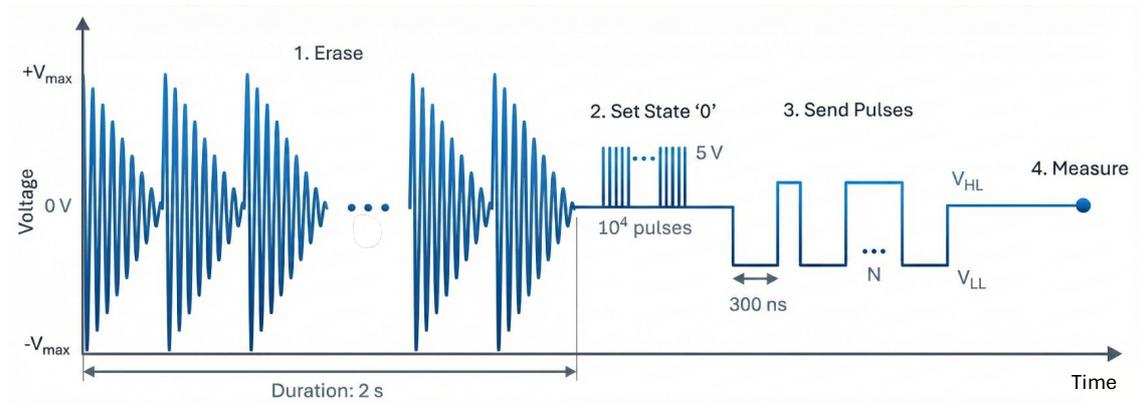

**Fig. S2. Electrical actuation and measurement protocol.** The characterization sequence consists of four steps. (1) Erase: A sinusoidal signal (1.3 MHz) amplitude-modulated with a down-ramp waveform (33.33 Hz) is applied for 2 s to randomize the ferroelectric domains and eliminate history effects. (2) Initialization: The device is poled into a saturated state (State '0') using a sequence of $10^4$ pulses (5 V). (3) Actuation: A train of N pulses (30 ns duration, 300 ns separation) with alternating high and low amplitudes is sent to induce partial switching. (4) Measurement: The voltage is set to 0 V (zero bias) to measure the non-volatile optical transmission.

**Supplementary Note 2.** Setup and automated graph-based characterization of the mesh.

### Experimental Setup

For the mesh measurements, the device was automatically characterized using a custom-built setup. A bank of power meters (EXFO LTB-12 mainframe with FTBx-1750 modules) was employed to monitor the optical power at the mesh output ports. The active components were driven by multi-channel voltage sources (Qontrol Ltd.).

### Automated Control Algorithm

The characterization routine relies on a directed graph abstraction of the mesh topology $G(V, E)$, where nodes represent the Mach-Zehnder Interferometers (MZIs) ports and edges represent the waveguide links[2]. The characterization process consists of three main steps (see Algorithm 1):

1. **Pathfinding:** For a given (Source, Target) pair of optical ports, the algorithm computes the shortest path using Dijkstra's algorithm. A geometric constraint is applied to avoid physical bends exceeding a maximum angle ($\theta_{max} = 120°$), ensuring no return by the same MZI (although other methods such as directed graphs can be used).
2. **Path Optimization:** Once the path is defined, the voltages of the MZIs along the path are optimized simultaneously using Particle Swarm Optimization (PSO) to maximize the optical power at the target port.
3. **Crosstalk Suppression:** To minimize crosstalk, the algorithm identifies the "adjacent" MZIs (nodes sharing a coupler with the active path but not part of it). The main path is temporarily detuned to induce leakage, and the adjacent MZIs are then optimized via PSO to minimize the detected power at the output, effectively blocking alternative paths.
4. **Individual Component Characterization:** With the routing configuration fully established, the system performs a sequential sweep of components. Each MZI along the path is individually swept across the voltage range 0-12 V, while keeping the rest of the path and adjacent MZIs at the optimized state. The resulting transmission curves are stored in a dataset, and the specific MZI is returned to its optimal voltage before proceeding to the next component.

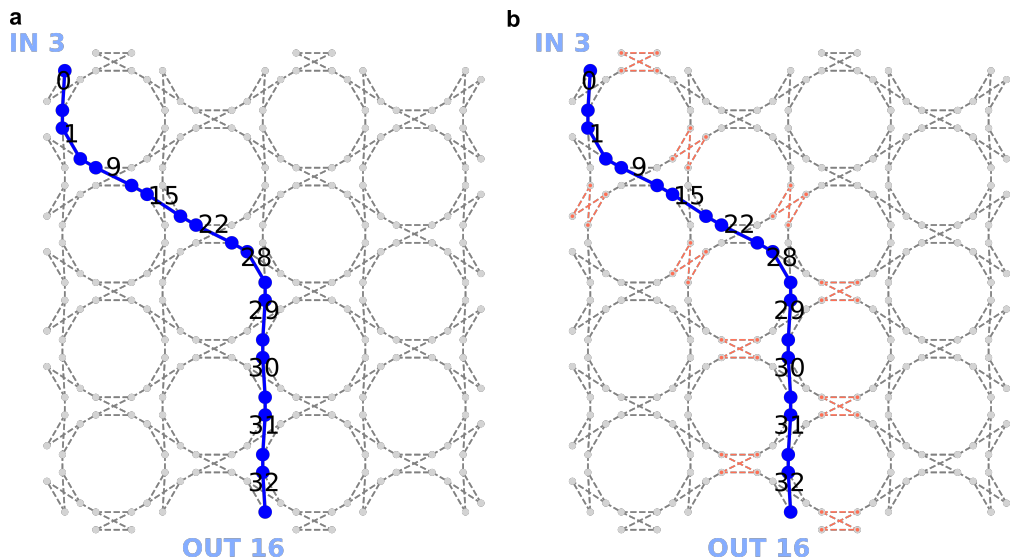

**Fig. S3. Graph-based routing and crosstalk suppression strategy. a**, Graph representation of the mesh topology showing the calculated shortest path (solid blue line) connecting Input 3 to Output 16. The labelled nodes correspond to the active MZIs (e.g., 0, 1, 9, 15...) whose voltages are optimized using PSO to maximize transmission. **b**, Visualization of the leakage suppression stage. The red dashed links indicate the adjacent MZIs and ports identified by the algorithm. These specific nodes are actively nulled to isolate the main path, minimizing optical crosstalk.

```
Input: Graph G, InputPort S, OutputPort T, MaxAngle θ_max
Output: Characterized Dataset for path S → T

1:  // PHASE 1: Topology Analysis & Pathfinding
2:  Initialize instruments (Laser, PowerMeter, VoltageSources)
3:  // Find shortest valid path using Dijkstra with angle constraint
4:  Path, MZIs_Active ← Pathfinder(G, S, T, θ_max)
5:  If Path is empty then Return Error

6:  // PHASE 2: Main Path Optimization
7:  Define Objective_Max(v):
8:      Apply voltages v to MZIs_Active
9:      Return -ReadPower(T) // Negative for minimization solver
10: v_path ← PSO_Optimize(Objective_Max, bounds=[0, V_π], iterations=60)
11: ApplyVoltage(MZIs_Active, v_path)

12: // PHASE 3: Leakage (Null Adjacent Nodes)
13: // Temporarily detune main path to force light into adjacent nodes
14: v_detuned ← v_path ± δ_leak
15: ApplyVoltage(MZIs_Active, v_detuned)
16: MZIs_Adj ← FindAdjacentNodes(G, Path)
17: Define Objective_Min(v):
18:     Apply voltages v to MZIs_Adj
19:     Return ReadPower(T) // Minimize leakage power
20: v_adj ← PSO_Optimize(Objective_Min, bounds=[0, V_π], iterations=60)
21: ApplyVoltage(MZIs_Adj, v_adj)
22: ApplyVoltage(MZIs_Active, v_path) // Restore optimal path

23: // PHASE 4: Sequential Sweep and Data Collection
24: Initialize Dataset D
25: For each mzi in MZIs_Active do:
26:     // Sweep current MZI from 0 to 12V
27:     SweepVoltage(mzi, range=[0, 12.0])
28:     Curve_Data ← RecordPower(T)
29:     Store Curve_Data in D
30:     // Restore optimal voltage for this MZI
31:     ApplyVoltage(mzi, v_path[mzi])
32: End For
33: Return D
```

**Algorithm 1.** Pseudo-code for the automated mesh characterization algorithm.

**Supplementary Note 3.** Extended experimental results: reconfigurable filtering.

Complementing the representative filter demonstrations discussed in the main text, this section presents the complete measurements of the synthesized optical filters, extending the analysis to configurations with longer cavity lengths and higher delays.

**Optical Ring Resonators (IIR Filters)**

In addition to the cavities discussed in the main text ($L_r = 6L_{PUC}$ and $10L_{PUC}$), we implemented a larger ring topology. Figure S4 summarizes the three synthesized topologies. The extended configuration (Ring 3) utilizes a cavity length of $14L_{PUC}$, achieving a reduced Free Spectral Range (FSR) of 44.5 pm.

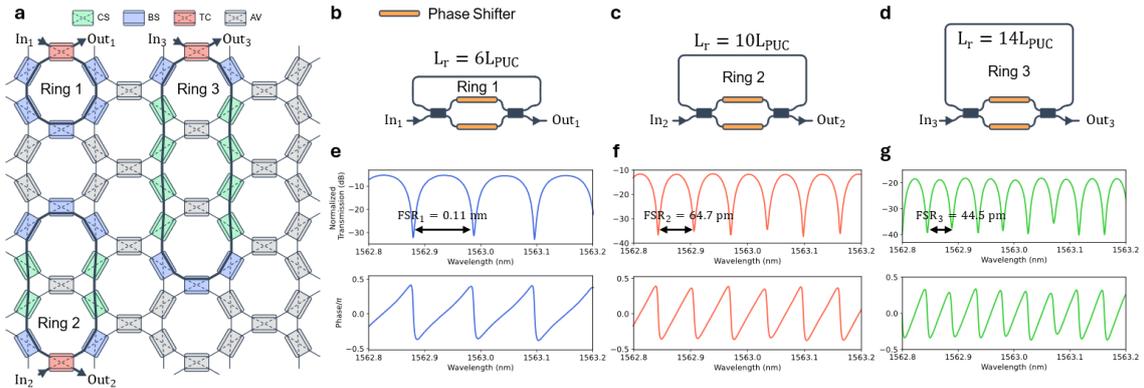

**Fig. S4. Extended characterization of Optical Ring Resonators (ORRs). a**, Schematic representation of the photonic mesh configured to support three distinct ring topologies. **b–d**, Block diagrams of the synthesized cavities. **e–g**, Measured normalized transmission spectra (top) and phase response (bottom).

**Unbalanced Mach-Zehnder Interferometers (FIR Filters)**

Similarly, the main text introduced finite impulse response filters with path differences of $\Delta L = 2L_{PUC}$ and $4L_{PUC}$. Here, we extend these measurements to a high-order delay line configuration with $\Delta L = 8L_{PUC}$ (Fig. S5). This configuration (UMZI 3) yields the finest spectral resolution among the FIR filters tested, with a measured FSR of 78.1 pm.

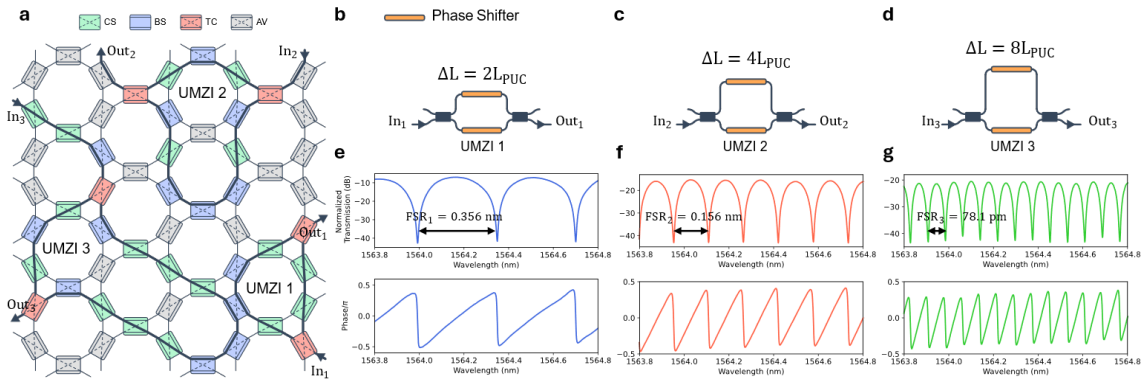

**Fig. S5. Extended characterization of Unbalanced Mach-Zehnder Interferometers (UMZIs). a**, Mesh topology showing the routing for three UMZIs with increasing delay lines. **b–d**, Diagrams illustrating the path length differences. **e–g**, Experimental spectral characterization. The plots display the transmission (top) and phase (bottom).

**Supplementary Note 4.** Thermal sensitivity of the barium titanate phase shifters.

To ensure the stability of the electro-optic reconfiguration, we characterized the thermal sensitivity of the Barium Titanate (BTO) phase shifters. A representative MZI within the mesh was isolated and characterized under varying thermal conditions using a Thermo-Electric Cooler (TEC) to precisely control the chip temperature. We performed voltage sweeps from 0 to 12 V at three distinct temperature setpoints: 20ºC, 25ºC, and 30ºC. As shown in Fig. S6, deviations in the operating temperature induce a shift in the phase response, altering the bias voltage required to achieve destructive interference. The voltage deviation between 25º and 20ºC is 0.45 V and between 25º and 30ºC is 0.24 V. That is why it is necessary to maintain the same temperature during characterization and measurements. Notably, the measurement at 25ºC (orange curve) exhibited the deepest resonance dip, achieving 42 dB of rejection. Conversely, deviations to 25ºC and 30ºC resulted in a degradation of the ER and a spectral shift. Consequently, a constant temperature of 25ºC was selected and actively maintained via the TEC for all the experiments presented in this work.

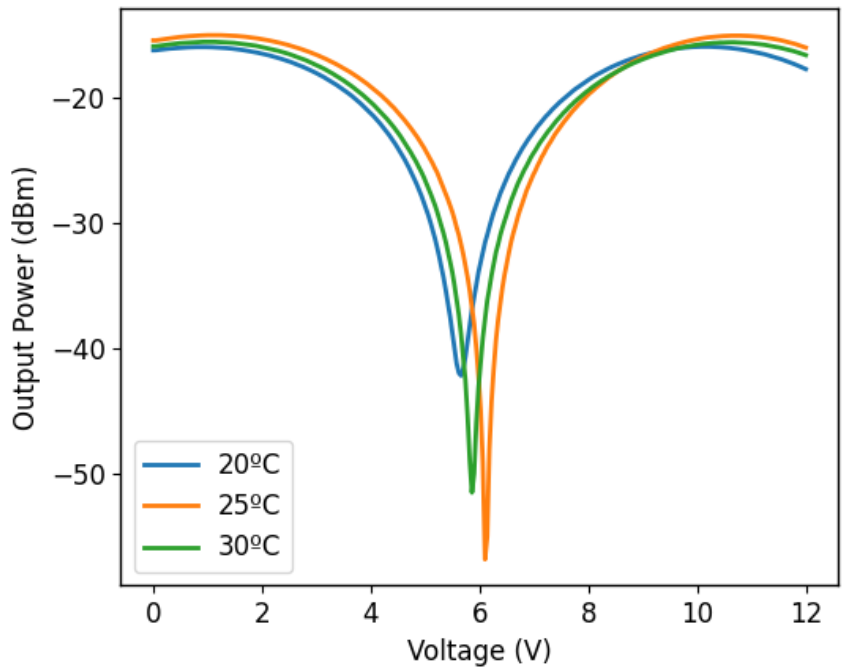

**Fig. S6.** Thermal dependence of a MZI with BTO phase shifters.

**Supplementary Note 5.** Fabricated silicon waveguide.

The photonic circuitry was processed on 200 mm base wafers fabricated at CEA-Leti using a silicon-on-insulator (SOI) platform featuring a 3 µm thick buried oxide (BOX) layer. While the platform is nominally based on a 220 nm silicon layer, characterization of the actual fabricated structures was verified.

To prepare a surface suitable for the heterogeneous integration of BTO, a planarized top cladding was required. The targeted top oxide thickness was 350 nm, which was achieved through a sequence of oxide deposition followed by a planarization step using Chemical-Mechanical Polishing (CMP).

Final validation was performed using Scanning Electron Microscopy (SEM) on a cross-section (Fig. S7). The imaging confirmed a robust rectangular waveguide profile with a core width of approximately 512.6 nm and a height of 242.5 nm. Furthermore, the critical top oxide thickness was measured at 365.7 nm. This value slightly exceeds the nominal target but ensures a sufficient planarization margin for the subsequent direct wafer bonding of the BTO film.

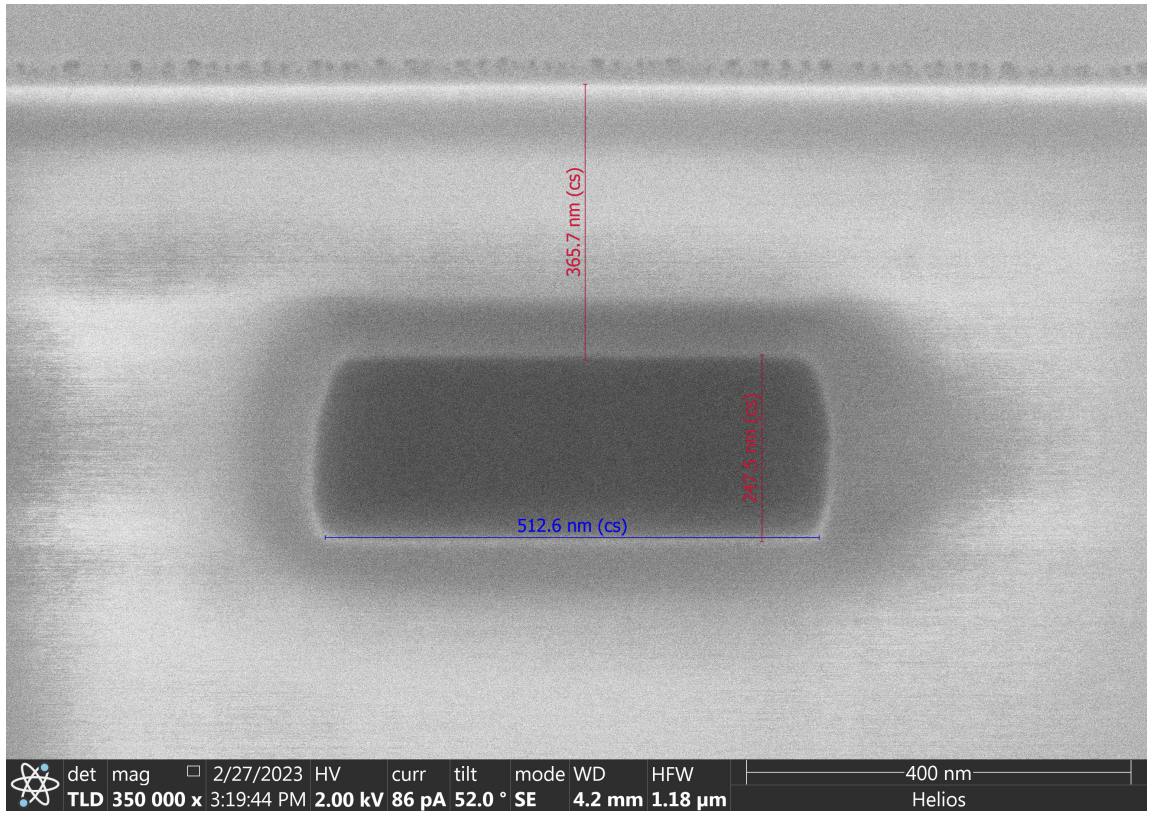

**Fig. S7.** Cross-sectional SEM analysis of the fabricated silicon waveguide.